\documentclass[11pt]{article}
\usepackage{amsmath, amssymb}
\usepackage{slashed}
\usepackage{verbatim}
\usepackage{latexsym}
\usepackage{euscript}
\usepackage{amsfonts}
\usepackage{verbatim}
\usepackage[]{array}
\usepackage[]{mathrsfs}
\usepackage{graphicx}
\usepackage{amsmath}
\usepackage{amssymb}
\usepackage{amsxtra}
\usepackage{color}
\def\be{\begin{eqnarray}}
\def\ee{\end{eqnarray}}
\def\0{\nonumber}

\def\bfh{{\bf h}}

\def\bfd{{\bf d}}
\def\bfD{{\bf D}}
\usepackage{slashed}
\usepackage{verbatim}
\usepackage{latexsym}\usepackage{color}
\usepackage{euscript}

\newcommand\ER{\EuScript{R}}

\newcommand\EA{\EuScript{A}}
\newcommand\ED{\EuScript{D}}
\normalfont\large
\begin{document}
\begin{flushright}
SISSA/26/2020/FISI
\end{flushright}
\vskip 2cm
\begin{center}
{\LARGE {\bf Supersymmetric HS Yang-Mills-like models}}

\vskip 1cm

{\large  L.~Bonora$^{a}$, S.~Giaccari$^{b}$,\\
\textit{${}^{a}$ International School for Advanced Studies (SISSA),\\Via
Bonomea 265, 34136 Trieste, Italy, and INFN, Sezione di
Trieste }\\ 
\textit{${}^{b}$ Department of Sciences,
Holon Institute of Technology (HIT),\\
52 Golomb St., Holon 5810201, Israel}
}
\vskip 1cm

\end{center}

 \vskip 1cm {\bf Abstract.}
We introduce the supersymmetric version of YM-like theories with infinitely many spin fields in 4 dimension. The construction is carried out via the superfield method. The surprising feature of these models is that they describe in particular gauge and gravity in a supersymmetric form with no need of supergravity. 

\vskip 1cm

\begin{center}

{\tt Email: bonora@sissa.it,stefanog@hit.ac.il}

\end{center}

\vskip 1cm

\section{Introduction}

The unification of gravity with the other fundamental forces in nature has always been a major and still unresolved challenge for theoretical physics. Of course superstring theories constitute the most authoritative candidate. But a teasing question is not easy to be ignored: are they the only possibility? In other words, is it not possible to construct a consistent {(i.e. unitary and renormalizable)} field theory that describes all interactions without resorting to string theory? Nowadays there is enough evidence that a theory describing all the fundamental interactions needs an infinite number of fields of increasing spins. This kind of considerations is at the origin of many attempts, which go under the name of higher spin (HS) { gauge} theories. {
{ Whereas the free spectrum of massless HS particles is by now well-understood ~\cite{Majorana:1968zz,Dirac:1936tg,Fierz:1939zz,Pauli:1939xp,Wigner:1939cj,Fronsdal:1978rb,Francia:2002aa,Francia:2007qt,Francia:2007ee,Campoleoni:2012th}, the construction of consistent interactions has come out as a much more challenging problem, a fundamental breaktrough being the discovery of Vasiliev's equations  ~\cite{Vasiliev:1990en,Vasiliev:1992av,Vasiliev:1995dn}.} { Unfortunately, no action is known for these theories, so that, in particular, their quantization is hindered. This dicovery was all the more remarkable since the constraints in constructing such theories are very severe.  There is in fact a wide
 literature where various constraints
(no-go theorems) have been produced for HS gauge theories on a flat Minkowski space (see e.g. ~\cite{Metsaev:1991mt,Ponomarev:2016lrm,Taronna:2017wbx,Roiban:2017iqg} and  \cite{Bekaert} and references within), and even on  anti-de Sitter (AdS) spaces ~\cite{Sleight:2017pcz}.}} For instance, there seems to be no hope for massless local theories and even for perturbatively local theories (that is, for theories with infinitely many interacting fields, but with finitely many interaction terms at each derivative order), and, no doubt, a solution, if any, must be searched in the realm of non-local theories. However once we decide to abandon the safe ground of local field theories with a finite number of fields, the number of possibilities grows beyond control and the true issue becomes how to tackle the problem of non-locality in an organized and manageable way.

In this spirit, in \cite{HSYM} a new type of massless higher spin theories on flat Minkowski spacetime was proposed. They were called HS YM-like models because the form of their actions is inspired by Yang-Mills theories and the lowest spin field is clearly a Yang-Mills field. {Inspired by the Weyl-Wigner quantization method,} the fields of the theories, $h_a(x,u)$, are local functions in phase space. This means that they  can be expanded in powers of the momentum  $u_\mu$ conjugate to the coordinate $x^\mu$. The expansion coefficients are ordinary local HS fields of which one can easily extract the equations of motion. The latter contain an infinite number of local terms of the coefficient fields and their derivatives, but the number of terms at each derivative order is finite. The main characteristic of the HS YM-like models is that they admit a gravitational interpretation. For the second field in the $u$ expansion has the correct quantum numbers of a frame field with the expected transformation properties. The action is invariant under gauge transformations whose parameters (local in the phase space) can be expanded in powers of $u$. As it turns out the lowest order term in such expansion represents a U(1) gauge transformation, while the next one corresponds to diffeomorphisms. The frame-like component of $h_a$ transforms correctly under them. However a full gravitational interpretation requires also invariance under local Lorentz transformations, while the action of the HS YM-like models does not  explicitly possess it. However, in \cite{HSYM} it was shown that such an invariance is hidden and can be easily made manifest by redefining the action in a suitable way. In summary, the gravitational interpretation  of the HS YM-like models is legitimate.

Ref. \cite{HSYM} contains also a proposal for quantization, in which the gauge freedom has to be fixed in order to extract the true physical degrees of freedom.  This is a complex operation which leads to an overall redefinition of the action, from where it appears that the theory is characterized by an intrinsic non-locality, although in a controlled and hopefully manageable way. The present paper will not continue this line of research, but rather examine another aspect the the HS YM-like theories. We want to show here that, in 4D, they possess a supersymmetric extension. {{ This is in line with a large amount of research developed over the years both to understand the ${\cal N}=1$ HS free spectrum ~\cite{Kuzenko:1993jp,Kuzenko:1993jq,Kuzenko:1994dm,Gates:2013rka,Gates:2013ska} and  to construct consistent interactions ~\cite{Buchbinder:2017nuc,Hutomo:2017phh,Hutomo:2017nce,Koutrolikos:2017qkx,Buchbinder:2018wwg,Buchbinder:2018wzq,Buchbinder:2018gle,Buchbinder:2018nkp,Gates:2019cnl}.} The supersymmetric version of the HS YM-like models} is easily implemented via the superfield formalism. The final action we will find is therefore supersymmetric and, at the same time, admits a gravitational interpretation. Therefore we have gravity and supersymmetry. The surprising aspect is that  supersymmetry is rigid, while at the same time, invariance under diffeomorphisms and local Lorentz transformations is guaranteed. In other words, there is no apparent need of and no room for supergravity. 

The paper is organized as follows. In section 2 we review the HS YM-like models. Sec. 3 is devoted to the supersymmetrization via the supermanifold formalism. In sec. 4 we show that the action is also local Lorentz invariant. Sec. 5 contains our conclusions.

\section{Review of HS-YM}

In this section we review the ordinary (non-supersymmetric) higher-spin Yang-Mills models. They are expressed in terms of master fields $h_a(x,u) $, which are local in the phase space $(x,u)$, with $[x^\mu, u_\nu]=i\hbar \delta^\mu_\nu$ ($\hbar$ will be set to the value 1). 
The master field can be expanded in powers of $u$,
\be
h_a(x,u) &=& \sum_{n=0}^\infty \frac 1{n!} h_a^{\mu_1,\ldots,\mu_n}(x) u_{\mu_1}\ldots u_{\mu_n}\0\\
&=&A_a(x) +\chi_a^\mu(x) u_\mu 
+ \frac 12 b_a^{\mu\nu}(x) u_\mu u_\nu+\frac 16 c_a^{\mu\nu\lambda}(x) u_\mu
u_\nu u_\lambda + \ldots\label{haunn}
\ee
where $ h_a^{\mu_1,\ldots,\mu_n}(x)$ are ordinary tensor fields, symmetric in $\mu_1,\ldots,\mu_n$. In the expansion \eqref{haunn} the indices $\mu_1,\ldots,\mu_n$ are upper (contravariant)
Lorentz indices, $\mu_i=0,\ldots,d-1$. The index $a$ is also a vector index, but is different in nature, it will be interpreted as a flat index and $h_a$ will be referred to as a {\it frame-like master field}\footnote{{The name frame-like field is a somewhat of an abuse of language, for $h_a$ contains only the fluctuating part of the frame field. Our (teleparallel) framework is in fact characterized by a splitting of the frame field into an inertial and a dynamical part. The two pieces are always separate and do not form a unique field like in ordinary gravity theories. This issue will be clarified in section 4.}}

Of course when the background metric {is flat} all indices are on the same footing, but
writing in this way leads to the correct interpretation.

The master field $h^a(x,u)$ may undergo gauge transformations
\be
\delta_\varepsilon h_a(x,u) = \partial_a^x 
\varepsilon(x,u)-i [h_a(x,u) \stackrel{\ast}{,} \varepsilon(x,u)] 
\equiv {\cal D}^{\ast x}_a  \varepsilon(x,u),
\label{deltahxp}
\ee
where we introduced the covariant derivative
\be
{\cal D}^{\ast x}_a = \partial_a^x- i  [h_a(x,u) \stackrel{\ast}{,}\quad].\0
\ee 
The $\ast$ product is the Moyal product, defined by
\be
f(x,u)\ast g(x,u) = f(x,u) e^{\frac i2 \left( \stackrel{\leftarrow}{\partial}_x \stackrel{\rightarrow}{\partial}_u - \stackrel{\leftarrow}{\partial}_u\stackrel{\rightarrow}{\partial}_x\right)}g(x,u) \label{Moyal}
\ee

In order to stress the difference between the index $a$ and the others, one can imitate the ordinary gauge theories using the compact notation  ${\bf d}= \partial_a\, dx^a, {\bf h}= h_a dx^a$ ($x^a$ are coordinates in the tangent spacetime) and  writing \eqref{deltahxp} as
\be
\delta_\varepsilon {\bf h}(x,u) ={\bf d} 
\varepsilon(x,u)-i [{\bf h} (x,u) \stackrel{\ast}{,} \varepsilon(x,u)] \equiv 
\bfD \varepsilon (x,u),\label{deltahxpbf}
\ee
 where it is understood that 
\be
  [{\bf h} (x,u) \stackrel{\ast}{,} \varepsilon(x,u)]=  [h_a (x,u)
\stackrel{\ast}{,} \varepsilon(x,u)]dx^a,\0
\ee
Next, one can introduce the curvature notation
\be
{\bf G} = {\bf d} {\bf h}
 -\frac i 2 [ {\bf h}\stackrel {\ast}{,}{\bf h}],\label{curv1} 
\ee
with the transformation property
\be
\delta_\varepsilon {\bf G} = -i [{\bf G}  \stackrel {\ast}{,}\varepsilon]
\label{deltaG}
\ee

The action functionals we will consider are  integrated polynomials of $\bf G$ or of
its components $G_{ab}$. In order to exploit the transformation property
\eqref{deltaG} in the construction we need the `trace property', analogous to
the trace of polynomials of Lie algebra generators in ordinary non-Abelian gauge
theories.  In the HS context such an object is 
\be
\langle\!\langle f\ast g\rangle\!\rangle\equiv \int d^dx \int \frac
{d^du}{(2\pi)^d}
f(x,u)\ast g(x,u) = \int d^dx \int \frac {d^du}{(2\pi)^d} f(x,u) g(x,u)=
\langle\!\langle g\ast f\rangle\!\rangle  \label{trace}
\ee
From this, plus associativity, it follows that
\be
&&\langle\!\langle f_1 \ast f_2\ast \ldots \ast f_n\rangle\!\rangle=
\langle\!\langle f_1 \ast (f_2\ast \ldots \ast f_n)\rangle\!\rangle\0\\
&&=(-1)^{\epsilon_1(\epsilon_2+\ldots+\epsilon_n)} \langle\!\langle  (f_2\ast
\ldots \ast f_n)\ast
f_1\rangle\!\rangle=(-1)^{\epsilon_1(\epsilon_2+\ldots+\epsilon_n)} 
\langle\!\langle  f_2\ast \ldots \ast f_n\ast f_1\rangle\!\rangle\label{cycl}
\ee
where $\epsilon_i$ is the Grassmann degree of $f_i$. In particular
\be
\langle\!\langle [f_1 \stackrel{\ast}{,} f_2\ast \ldots \ast
f_n\}\rangle\!\rangle=0\label{comm0}
\ee
where $[\quad  \stackrel{\ast}{,}\quad\}$ is the $\ast$-commutator or
anti-commutator, as appropriate.

This property holds also when the $f_i$ are valued in a (finite dimensional) Lie algebra, provided
the symbol
$\langle\!\langle   \quad  \rangle\!\rangle$ includes also the trace over the
Lie algebra generators.

\subsection{HS Yang-Mills action}

The curvature components, see \eqref{curv1}, are
\be
G_{ab}= \partial_a h_b - \partial_b h_a -i [h_a \stackrel{\ast}{,} h_b ]
\label{Gab}
\ee
Their transformation rule is
\be
\delta_\varepsilon G_{ab}=-i [G_{ab}\stackrel{\ast}{,}
\varepsilon]\label{deltaGab}
\ee
If we consider the functional $\langle\!\langle G^{ab} \ast G_{ab} \rangle\!\rangle$, it follows from the above that
\be 
\delta_\varepsilon \langle\!\langle G^{ab} \ast G_{ab} \rangle\!\rangle = -i
\langle\!\langle 
   G^{ab} \ast G_{ab} \ast\varepsilon -\varepsilon \ast  G^{ab} \ast G_{ab}
\rangle\!\rangle=0 \label{YMinvariance}
\ee
Therefore 
\be
{\cal YM}({\bfh})=- \frac 1{4 g^2}\langle\!\langle G^{ab} \ast G_{ab}
\rangle\!\rangle\label{YMh}
\ee
is invariant
under HS gauge transformations and it is a well defined  primitive functional
in any dimension.

\subsection{The non-Abelian case}

All that has been done for the Abelian $U(1)$ case up to now can be repeated for
the non-Abelian case with minor changes. One simply assumes that the master field $h_a$ belongs to the adjoint representation of a non-Abelian Lie algebra with
generators $T^\alpha$. 
\be
{\mathsf h}= {\mathsf h}^\alpha T^\alpha, \quad\quad {\mathsf h}^\alpha =
h^\alpha_a
dx^a\label{halpha}
\ee
where summation over $\alpha$ is understood.  The HS gauge parameter is now
\be
{\mathsf e} (x,u) = \varepsilon^\alpha(x,u) T^\alpha\label{epsilonnA}
\ee
and the transformation of $ {\mathsf h} (x,u) $  
\be
\delta_{\mathsf e} {\mathsf h} (x,u) = \bfd^x 
	{\mathsf e}(x,u)-i [{\mathsf h}(x,u) \stackrel{\ast}{,}{\mathsf e}
(x,u)] ,
\label{deltahxpnA}
\ee
if the generators $T^\alpha$ are anti-hermitean. In this case the curvature is
\be
{\mathsf G} = {\bf d} {\mathsf h}
 -\frac i 2 [ {\mathsf h}\stackrel {\ast}{,}{\mathsf h}]\label{curv2}
\ee
The $\ast$-commutator includes now also the Lie algebra commutator. Of course we
have, in particular,
\be
\delta_{\mathsf e} {\mathsf G} (x,u)= -i [{\mathsf G}(x,u)
\stackrel{\ast}{,}{\mathsf e} (x,u)]
\ee
Everything works as before provided the symbol $\langle\! \langle \quad
\rangle\!\rangle$ comprises also the trace over the Lie algebra generators.
In particular
\be
{\cal YM}({\mathsf h})=- \frac 1{4 g^2}\langle\!\langle {\mathsf G}^{ab} \ast
{\mathsf G}_{ab}
\rangle\!\rangle\label{YMhnonAb}
\ee
is invariant under the HS non-Abelian gauge transformations and it is a well defined 
functional in any dimension.

The expressions \eqref{YMh} { and \eqref{YMhnonAb} do not have the usual form of field theory actions, }
because they are integrals over the phase space of the point
 particle with coordinate $x^a$. In particular their dimension is 
not the one of an action. One should divide it by a factor ${\cal V}_u$  proportional to the
integration volume over the momentum space. Since it does not play any role at the classical level, for the sake of simplicity, we disregard this factor. 
Nevertheless we can extract from \eqref{YMh} covariant
eom's by  taking the variation with respect to $\bfh$. In other words we assume 
that the action principle holds for fields defined in the phase space. This has been justified in 
\cite{HSYM}.

\subsection{Covariant YM-type eom's}
\label{s:eoms}

From\eqref{YMh}  we get the following eom:
\be
\partial_b G^{ab} -i [h_b\stackrel{\ast}{,} G^{ab}]\equiv{\cal D}_b^\ast
G^{ab}=0\label{YMeom}
\ee
which is, by construction, covariant under the HS gauge transformation
\be
\delta_\varepsilon\left( {\cal D}_b^\ast G^{ab}\right)= -i[ {\cal D}_b^\ast
G^{ab},\varepsilon]\label{covYMeom}
\ee
In components this equation splits into an infinite set according to the powers
of $u$. Let us expand  $G_{ab}$ in powers of $u$. We have
\be
G_{ab}&=&  F_{ab}+ X_{ab}^\mu u_\mu +\frac 12 B_{ab}^{\mu\nu} u_\mu u_\nu
+\frac 16 C_{ab}^{\mu\nu\lambda}u_\mu u_\nu u_\lambda
+ \frac 1{4!}  D_{ab}^{\mu\nu\lambda\rho}u_\mu u_\nu u_\lambda u_\rho
+\ldots\label{Gabcomp}
\ee
Of course $F_{ab}, X^\mu_{ab},...$ are local expressions of the component fields of $h_a$.

More explicitly, for instance, the first eom is
\be
0&=& \square A_b -\partial_b \partial \!\cdot\!A +  \left(
\partial_\sigma \partial \!\cdot\!A \chi_b^\sigma
+\partial_\sigma A^a \partial_a \chi_b^\sigma - \partial_\sigma \partial^a A_b
\chi_a^\sigma 
- \partial_\sigma  A_b \partial
\!\cdot\!\chi^\sigma\right)\label{squareFabexpl}\\
&&+ \partial_\sigma A^a \left( \partial_a \chi_b^\sigma - \partial_b
\chi_a^\sigma
+\frac 12 \left( \partial_\lambda A_a b_b^{\lambda\sigma} -\partial_\lambda A_b
b_a^{\lambda\sigma} +
\partial_\lambda \chi_a^\sigma \chi_b^\lambda 
- \partial_\lambda \chi_b^\sigma \chi_a^\lambda\right)\right) \0\\
&& - \chi_a^\sigma \left( \partial_\sigma \partial^a A_b -
\partial_\sigma \partial_b A^a 
+ \frac 12 \left(\partial_\sigma \partial_\lambda A^a \chi^\lambda_b 
+\partial_\lambda A^a \partial_\sigma \chi^\lambda_b-  
\partial_\sigma \partial_\lambda A_b \chi^{a\lambda} 
-\partial_\lambda A_b \partial_\sigma \chi^{a\lambda}\right)\right) 
\0\\
&&+\ldots\ldots\0
\ee

\be
\square \chi_a^\mu- \partial_a \partial^b \chi_b^\mu&=&  
\partial^b( \partial_\sigma A_a \,b_b^{\sigma \mu}- \partial_\sigma A_b
\,b_a^{\sigma \mu}+\partial_\sigma \chi_a^\mu \chi_b^\sigma- \partial_\sigma
\chi_b^\mu \chi_a^\sigma)\label{eomchi}\\
&&+ \partial_\tau A^b \partial_a b_b^{\mu\tau} -  \partial_\tau A^b \partial_b
b_a^{\mu\tau}+ 
\partial_\tau \chi^{b\mu} \partial_a\chi_b^\tau - \partial_\tau \chi^{b\mu}
\partial_b\chi_a^\tau\0\\
&&- \partial_\tau\partial_a A_b\, b^{b\tau\mu}+\partial_\tau\partial_b A_a\,
b^{b\tau\mu}-  \partial_\tau\partial_a \chi_b^\mu
\chi^{b\tau}+\partial_\tau\partial_b \chi_a^\mu \chi^{b\tau} +\ldots
\ee
where the ellipses in the RHS refer to terms containing {three or more}
derivatives.  As is clear from \eqref{squareFabexpl}, for instance, the above
eom's 
are characterized by the fact that at each order, defined by the number of
derivatives, there is a finite number of terms. {We  call a theory with this characteristic {\it
perturbatively local}
}.  

\subsection{Gauge transformations}

In this subsection we examine more in detail the gauge transformation \eqref{deltahxp} and propose an interpretation of the lowest spin fields. Let us expand the master gauge parameter $\varepsilon(x,u)$
\be
\varepsilon(x,u)&=& \epsilon(x) +\xi^\mu(x) u_\mu+\frac 12
\Lambda^{\mu\nu}(x)u_\mu
u_\nu+\frac 1{3!} \Sigma^{\mu\nu\lambda}(x)u_\mu
u_\nu u_\lambda+\ldots\label{epsxu}
\ee
and consider the first few terms in the transformation law of the lowest spin fields ordered in such a way that component fields and gauge parameters are infinitesimals of the same order. To lowest order the transformation \eqref{deltahxp} reads
\be
&& \delta^{(0)} A_a= \partial_a \epsilon\0\\
&&\delta^{(0)} \chi_a^{\nu} = \partial_a \xi^\nu  \0\\
&&\delta^{(0)}b_a{}^{\nu\lambda} = \partial_a\Lambda^{\nu\lambda}
\label{deltaAhb}
\ee

To first order we have
\be
\delta^{(1)} A_a &=& \xi\!\cdot\!\partial A_a - \partial_\rho \epsilon
\,\chi_a^{\rho} \label{delta1Ahb}\\
\delta^{(1)} \chi_a^{\nu} &=& \xi\!\cdot\!\partial \chi_a^\nu-\partial_\rho
\xi^\nu \chi_a^\rho 
+ \partial^\rho A_a \Lambda_{\rho}{}^\nu   - \partial_\lambda \epsilon
\,b_a{}^{\lambda\nu}\0\\
\delta^{(1)} b_a^{\nu\lambda} &=& \xi\!\cdot\!\partial b_a{}^{\nu\lambda}
-\partial_\rho \xi^\nu b_a{}^{\rho\lambda}- \partial_\rho \xi^\lambda 
b_a{}^{\rho\nu }+{\partial_\rho \chi_a^{\nu} \Lambda^{\rho\lambda}
+\partial_\rho
\chi_a^{\lambda} \Lambda^{\rho\nu}}
- \chi_a^{\rho} \partial_\rho \Lambda_{\nu\lambda} \0
\ee
The next orders contain three and higher derivatives.

These transformation properties allow us to associate the first two component fields of $h_a$ to an ordinary  U(1) gauge field  and to a vielbein. To see this
let us denote by {$\tilde A_a$ and $\tilde E_a^\mu = \delta_a^\mu -\tilde \chi_a^\mu$}
the standard gauge and vielbein fields. The standard gauge and diff
transformations, are
{ 
\be
\delta \tilde A_a&\equiv& \delta \left(\tilde E_a^\mu \tilde A_\mu\right)\equiv
\delta  \left((\delta_a^\mu -\tilde \chi_a^\mu) \tilde
A_\mu\right)\label{standardtransf}\\
&=&\left(-\partial_a \xi^\mu -\xi\!\cdot\!\partial \tilde \chi_a^\mu +\partial_\lambda \xi^\mu
\tilde \chi_a^\lambda\right) \tilde A_\mu+(\delta_a^\mu -\tilde
\chi_a^\mu)\left(\partial_\mu\epsilon + \xi\!\cdot\! \partial \tilde A_\mu + \tilde A_\lambda \partial_\mu \xi^\lambda\right) \0\\
&= &
\partial_a\epsilon + \xi\!\cdot\! \partial \tilde A_a- \tilde \chi_a^\mu
\partial_\mu\epsilon \0 
\ee
}
and
{
\be
\delta \tilde E_a^\mu \equiv  \delta  (\delta_a^\mu -\tilde \chi_a^\mu) =
\xi\!\cdot\! \partial\tilde e_a^\mu -\partial_\lambda \xi^\mu \tilde e_a^\lambda = -
\xi\!\cdot\!\partial \tilde \chi_a^\mu -\partial_a \xi^\mu +\partial_\lambda \xi^\mu
\tilde \chi_a^\lambda\label{deltaeamu}
\ee
}
so that
\be
\delta \tilde \chi_a^\mu= \xi\!\cdot\! \partial\tilde \chi_a^\mu +\partial_a \xi^\mu
-\partial_\lambda \xi^\mu \tilde \chi_a^\lambda\label{deltaeamu1}
\ee
where we have retained only the terms at most linear in the fields.

Now it is important to understand the derivative $\partial_a$ in eq.\eqref{deltahxp} and \eqref{deltaAhb} in the appropriate way:  the
derivative $\partial_a$ means $\partial_a = \delta_a^\mu \partial_\mu,$ not {
$ \partial_a = E_a^\mu \partial_\mu= \left(\delta_a^\mu
-\chi_a^\mu\right)\partial_\mu$}.
In fact the linear correction $ -\chi_a^\mu\partial_\mu$ is
contained in the term $ -i [h_a(x,u) \stackrel{\ast}{,} \varepsilon(x,u)]$, see
for instance the second term in the RHS of the first equation
\eqref{delta1Ahb}. 

From the above it is now immediate to make the identifications
\be
A_a= \tilde A_a, \quad\quad \chi_a^\mu = \tilde \chi_a^\mu\label{identAAee}
\ee

The transformations \eqref{deltaAhb}, \eqref{delta1Ahb}
allow us to interpret  $\chi_a^\mu$ as the fluctuation of the inverse 
vielbein, therefore the effective action may 
accommodate gravity. However a gravitational interpretation requires also that the frame field transforms under local Lorentz transformations. Therefore we expect that the master field $h_a$ transforms and the action be invariant under local Lorentz transformations. This is an important aspect of the theory, which, in \cite{HSYM}, has been proven to be implementable in HS YM-like theories. We will understand it for the moment, but we will resume this subject in section 5 in a more general supersymmetric setting. Therefore a HS YM-like theory admit a gravitational interpretation\footnote{{For the sake of clarity: we are not claming our theory is equivalent to Einstein-Hilbert gravity or, even, teleparallel gravity. Its physical significance needs still to be uncovered and can only come out of the analyis of the equations of motion and the interactions  involving an infinite  number of fields (not, for instance, from a simple truncation to spin 1 and 2).}}

It should be emphasized that this is an entirely new approach to gravity. The gauge
transformation \eqref{deltahxp} reproduces both ordinary U(1) gauge transformations and
diffeomorphisms, but the eom of $\chi_a^\mu$ are not quite the ordinary
metric equations of motion: the linear eom coincides with the ordinary one after gauge
fixing, but there is a huge difference with ordinary gravity because in the latter the
interaction terms are infinite and include all powers in the fluctuating field, while in the
\eqref{YMh} there are at most cubic interaction terms. 

The HS YM-like approach to gravity is not entirely new. An approach close to it (but involving only gravity) has been considered in the past under the name of  {\it teleparallel gravity}, see \cite{teleparallel} and section \ref{s:LLC} below.
\vskip 1cm

\subsection{Scalar and spinor master fields}

We can couple to the HS YM-like theories matter-type fields of any spin, for instance, 
a complex multi-boson field 
\be
\Phi(x,u)= \sum_{n=0}^\infty \frac 1{n!}\Phi^{\mu_1\mu_2\ldots \mu_n}(x)
 u_{\mu_1}u_{\mu_2}\ldots u_{\mu_n}\label{Phixu}
\ee
which, under a master gauge transformation \eqref{deltahxp} transforms like $
\delta_\varepsilon \Phi = i \varepsilon\ast\Phi$.
We define as well the covariant derivative $ \ED^\ast_a \Phi= \partial_a \Phi -i h_a \ast \Phi$, which has the property $\delta_\varepsilon \ED^\ast_a \Phi= i\,\varepsilon \ast \ED^\ast_a \Phi$.
It follows that the kinetic action term $
\frac 12 \langle\!\langle(\ED_\ast^a
\Phi)^\dagger\ast\ED^\ast_a \Phi\rangle\!\rangle$ and potential terms 
$ \langle\!\langle(\Phi^\dagger \ast \Phi)^n_\ast\rangle\!\rangle$ are
 gauge invariant.

\vskip 1cm

In a very similar manner we can introduce master spinor fields,
\be
\Psi(x,u) = \sum_{n=0}^\infty\frac 1{n!} \Psi_{(n)}^{\mu_1\ldots \mu_n}(x)
u_{\mu_1} \ldots u_{\mu_n}, \label{Psi}
\ee
where $\Psi_{(0)}$ is a Dirac field. Under HS gauge transformations it transforms according to
$\delta_\varepsilon \Psi = i \varepsilon\ast\Psi $, so we can define the covariant derivative
\be
\ED^\ast_a \Psi= \partial_a \Psi -i h_a \ast \Psi\label{covderPsi}
\ee
so that $\delta_\varepsilon (\ED^\ast_a \Psi)=i\varepsilon \ast (\ED^\ast_a
\Psi)$.
It is evident that the action
\be
S(\Psi,h) = \langle\!\langle \overline \Psi i\gamma^a \ED_a \Psi
\rangle\!\rangle
=  \langle\!\langle \overline \Psi \gamma^a\left(i\partial_a+ h_a \ast\right)
\Psi \rangle\!\rangle
\label{Spsih}
\ee
is invariant under the HS gauge transformations \eqref{deltahxp}. 
\vskip 1cm
{This completes the presentation of classical HS YM models. Before passing to their supersymmetric version we would like to make a short comment on their physical content. Any HS YM models contains plenty of unphysical states. In \cite{HSYM} it was shown how to deal with them. The analysis of the quadratic kinetic operators suggests that only the traceless transverse fields have a propagator, thus only the transverse among them are physical. It is always possible to define projectors into the traceless transverse representations of the Lorentz group. This means that, inserting these projectors inside the amplitudes defined with the usual Feynman diagram procedure, one obtains amplitudes where only physical states propagate, and unphysical states are excluded. It is possible to implement the restriction to traceless transverse fields at the Lagrangian level, although this leads to severe non-localities of the type $\frac 1{\square^n}$. The new non-local actions are invariant under two HS gauge symmetries, which account for the elimination of the unphysical degrees of freedom. The non-locality is however to be considered a gauge artifact, because it disappears if we return to the initial gauge-fixed theories.  }

Let us pass now to the supersymmetric version of HS YM models in 4d.

\section{Supersymmetrization}
\label{s:supersymmetrization}

The easiest way to supersymmetrize HS YM models is by means of the superfield formalism. Using the notation and conventions of \cite{BW} we introduce a (real) master vector superfield, 
\be
V(x,u,\theta, \bar\theta)&=& C(x,u) + i \theta \chi(x,u) -i \bar\theta\bar\chi(x,u) +i\theta\theta \left[ M(x,u)+iN(x,u)\right]\0\\
&&-i \bar\theta\bar\theta  \left[ M(x,u)-iN(x,u)\right]-\theta \sigma^a \bar \theta\,  V_a(x,u) +i \theta\theta \bar\theta\left[ \bar \lambda(x,u)+\frac i2 \bar\sigma^a \partial_a \chi(x,u)\right]\0 \\
&&-i \bar\theta\bar\theta \theta \left[ \lambda(x,u) +\frac i2 \sigma^a \partial_a\bar \chi(x,u)\right] +\frac 12\theta\theta \bar\theta \bar \theta\left[  D(x,u)+\frac 12 \square C(x,u)\right]\label{Vxu}
\ee
Let us briefly explain the notation. We use the two component formalism for spinors. A Dirac fermion is represented by
\be
\psi_D = \left(\begin{matrix} \lambda_\alpha \\ \bar \chi^{\dot\alpha}\end{matrix}\right),\quad\quad \alpha=1,2, \quad\quad \dot\alpha=\dot 1, \dot 2. \label{Diracpsi}
\ee
and a Majorana one by
\be
\psi_M= \left(\begin{matrix} \lambda_\alpha \\ \bar \lambda^{\dot\alpha}\end{matrix}\right),\quad\quad \alpha=1,2, \quad\quad \dot\alpha=\dot 1, \dot 2. \label{Majopsi}
\ee
In term of ordinary gamma matrices this formalism corresponds to the choice
\be
\gamma^a= \left( \begin{matrix} 0 & \sigma^a \\ \bar\sigma^a &0\end{matrix}\right),\quad\quad \gamma_5= \left( \begin{matrix} 1&0\\0&-1\end{matrix}\right)\label{gammamatrices}
\ee
where $a=(0,i)$ with $i-1,2,3$ is a flat four-index.
\be
\sigma^a =(1, \sigma^i) \quad\quad \bar \sigma^a = (1,-\sigma^i), \quad\quad i=1,2,3\0
\ee
the $\sigma^i= \sigma_{\alpha\dot\alpha}{}^i$ being the Pauli matrices. In \eqref{Vxu} we 
use the compact notation
\be
\lambda\chi=\lambda^\alpha \chi_\alpha, \quad\quad \bar\lambda \bar\chi= \bar\lambda_{\dot\alpha}\bar \chi^{\dot \alpha}\label{lambdachi}
\ee
where $\lambda^\alpha =\varepsilon ^{\alpha\beta} \lambda_\beta$ and $\bar\lambda_{\dot\alpha}= \varepsilon_{\dot\alpha \dot\beta} \bar\lambda^{\dot\beta}$. $\varepsilon$ is the antisymmetric symbol with $\varepsilon^{12}=\varepsilon_{21}=1$, $\varepsilon_{12}=\varepsilon^{21}=-1$ and $\varepsilon_{11}=\varepsilon_{22}=0$.

Returning to \eqref{Vxu}, we can trade the master fields $C,M,N,\chi,\bar\chi$ with a gauge transformation and disregard them. This means making a gauge choice, the Wess-Zumino gauge. In this gauge $V$ takes the form
\be
V(x,u,\theta, \bar\theta)= -\theta \sigma^a \bar \theta V_a(x,u) +i \theta\theta \bar\theta \bar \lambda(x,u)-i \bar\theta\bar\theta \theta \lambda(x,u) +\frac 12 \theta\theta \bar\theta \bar \theta  D(x,u)\label{VxuWZ}
\ee
In the sequel $V_a(x,u)$ will be related with the symbol $h_a(x,u)$ of the previous section, while $\lambda(x,u)$ is its fermionic superpartner. $D(x,u)$ is a bosonic auxiliary master field.

This simplification is possible because of a supergauge symmetry. 
An HS supergauge transformation is introduced in this system by means of a master field parameter
$E(x,u,\theta,\bar\theta)$ and its conjugate\footnote{Hermitean conjugation implies in particular
\be
(\chi\lambda)^\dagger = (\chi^\alpha \lambda_\alpha)^\dagger = \bar \lambda_{\dot\alpha} \bar\chi^{\dot\alpha}= \bar \lambda \bar\chi =\bar\chi \bar\lambda\0
\ee}
 $E(x,u,\theta,\bar\theta)^\dagger$. A finite supergauge transformation is defined by
\be
e^V_\ast \to e^{-\frac i2 E^\dagger}_\ast\, \ast\, e^V_\ast\,\ast\, e^{\frac i2 E}_\ast\label{supergauge} 
\ee
For an infinitesimal $E$ we get the transformation
{
\be
\delta V&=& \frac{i}2(E-E^\dagger)+\frac{i}2\left[\frac V2 \stackrel{\ast}{,}( E+E^\dagger)\right]\label{infinitesimalE}\\
& & +\frac{i}6\left[\frac V2 \stackrel{\ast}{,}\left[ \frac V2 \stackrel{\ast}{,}( E+E^\dagger)\right]\right] + O(V^3)
\ee 
}
We assume $E$ to be chiral, i.e. to satisfy
\be
\bar D_{\dot \alpha} E=0,    \quad\quad \bar D_{\dot\alpha} = -\frac {\partial}{\partial \bar\theta^{\dot\alpha}}- i\theta^\alpha\sigma_{\alpha\dot\alpha}{}^a \partial_a\label{Dbaralpha}
\ee
Consequently $E^\dagger$ is antichiral
\be
 D_{\alpha} E^\dagger=0,    \quad\quad D_{\alpha} = \frac {\partial}{\partial \theta^{\alpha}}+ i  \sigma_{\alpha\dot\alpha}{}^a \bar\theta^{\dot\alpha} \partial_a
\label{Dalpha}
\ee
We recall that the superspace covariant derivatives satisfy the relations
\be
&&[D_\alpha, D_\beta]=0,\quad\quad [\bar D_{\dot\alpha}, \bar D_{\dot\beta}]=0,\0\\
&&[D_\alpha,  \bar D_{\dot\beta}]= -2i \sigma_{\alpha\dot\beta}{}^a \partial_a\label{DDbarDbarD}
\ee

The number of component fields in $E$ is limited by this condition, see \cite{BW}. The relevant combinations in  \eqref{infinitesimalE} are $E\pm E^\dagger$. Their expansion is
\be
E(x,u,\theta,\bar\theta)\!\!\! &\pm&\!\!\! E(x,u,\theta,\bar\theta)^\dagger= e(x,u)\pm e(x,u)^*+\sqrt{2}\left( \theta \psi(x,u) \pm \bar\theta\bar\psi(x,u)\right)\label{Exu}\\
&&+\theta\theta F(x,u)\pm \bar\theta\bar\theta  F(x,u)^*+i \theta \sigma^a \bar \theta\, \partial_a (e(x,u)\mp e(x,u)^*)\0\\
&&+\frac i{\sqrt{2}}\theta\theta\bar\theta  \bar\sigma^a \partial_a \psi(x,u)\pm \frac i{\sqrt{2}}\bar\theta\bar\theta\theta  \sigma^a \partial_a \bar\psi(x,u)+ \frac 14 \theta\theta \bar\theta \bar \theta \square( e(x,u)\pm e(x,u)^*)\0
\ee
Considering the {first term}
of the transformation \eqref{infinitesimalE} it  is easy to see that we can get rid of $C,M,N,\chi$ and $\bar\chi$ by using the components of $E-E^\dagger$ { except one}.
{ In the Wess-Zumino gauge  \eqref{infinitesimalE} becomes
\be
\delta V=\frac{i}2(E-E^\dagger)+\frac{i}2\left[\frac V2 \stackrel{\ast}{,}( E+E^\dagger)\right]\,,
\ee 
where 
\be
E(x,u,\theta,\bar\theta)=e^{i \theta \sigma^a \bar \theta\, \partial_a} \left(\frac{e(x,u)+e(x,u)^*}2\right)\,.
\ee
}
Comparing with \eqref{Vxu} we see that
\be
\delta V_a(x,u) ={\frac12} \partial_a\bigl{(}e(x,u)+e(x,u)^*\bigr{)}+\ldots \quad\quad \delta \lambda(x,u)=0+\ldots, \quad\quad \delta D(x,u)=0+\ldots \label{incomplete}
\ee 
Taking into account the nonlinear part of  \eqref{infinitesimalE} we recover the complete first order gauge transformations
\be
\delta V_a(x,u) &=& {\frac12}\partial_a\bigl{(}e(x,u)+e(x,u)^*\bigr{)}+\frac i2\bigl{ [ }V_a(x,u) \stackrel{\ast}{,}\bigl{(}{\frac{e(x,u)+e(x,u)^*}2}\bigr{)}\bigr{]}\label{deltaVaxu}\\
\delta \lambda(x,u) &=& \frac i2\bigl{ [ }\lambda(x,u) \stackrel{\ast}{,}\bigl{(}{\frac{e(x,u)+e(x,u)^*}2}\bigr{)}\bigr{]}\label{deltalambdaxu}\\
\delta D_a(x,u) &=&\frac i2\bigl{ [ }D_a(x,u) \stackrel{\ast}{,}\bigl{(}{\frac{e(x,u)+e(x,u)^*}2}\bigr{)}\bigr{]}\label{deltaDxu}
\ee
To make contact with HS YM we have to identify $e(x,u)+e(x,u)^*$ with {$-2 \varepsilon(x,u)$} and $V_a(x,u)$ with {$-2 h_a(x,u)$}.

Next, as usual, we define
\be
W_\alpha (x,u) = -\frac 14 \bar D\bar D \left(e^{-V(x,u)}_\ast\ast D_\alpha e^{V(x,u)}_\ast\right) \label{Walpha}
\ee
This master superfield is chiral: $\bar D_{\dot\alpha} W_\alpha (x,u)=0$ and under a supergauge transformation \eqref{supergauge} it transforms as
\be
W_\alpha\rightarrow {e^{-\frac{i}2E}_\ast \ast W_\alpha \ast e^{\frac{i}2E}_\ast }\label{Wtranf}
\ee
Similarly
\be
\bar W_{\dot\alpha} (x,u) = \frac 14  D D \left(e^{V(x,u)}_\ast\ast \bar D_{\dot \alpha} e^{-V(x,u)}_\ast\right) \label{Wbaralpha}
\ee
which is antichiral and transform as
\be
\bar W_{\dot\alpha}\rightarrow {e^{-\frac{i}2E^\dagger}_\ast \ast \bar W_{\dot\alpha} \ast e^{\frac{i}2E^\dagger}_\ast} \label{Wbartranf}
\ee

Let us introduce the 2x2 matrices
\be
\sigma_{ab\alpha}{}^\beta &=&\frac 14 \bigl{(} \sigma_{a\alpha \dot\alpha} \bar\sigma_b{}^{\dot\alpha \beta} - \sigma_{b\alpha \dot\alpha} \bar\sigma_a{}^{\dot\alpha \beta}\bigr{)}
 \0\\
{\bar\sigma_{ab}{}^{\dot\alpha}}{}_{\dot\beta}&=& \frac 14 \bigl{(} {\bar\sigma}_a{}^{{\dot\alpha} \alpha} \sigma_{b\alpha \dot\beta}-  {\bar\sigma}_b{}^{{\dot\alpha} \alpha} \sigma_{a\alpha \dot\beta}\big{)}\0
\ee
Notice that
\be
\frac 14 (\gamma_a\gamma_b-\gamma_b\gamma_a) = \left( \begin{matrix} \sigma_{ab} &0\\
0&\bar\sigma_{ab} \end{matrix}\right)\0
\ee
are the generators of the Lorentz algebra.

Among the lowest order terms of $W_\alpha $ we have
\be
W_\alpha = ... -\frac i 4 {\sigma^{ab}}_{\alpha}{}^\beta \theta_\beta W_{ab}(x,u) +\ldots
\ee
where
{
\be
W_{ab}(x,u) = 2\left(\partial_a V_b(x,u) - \partial_b V_a(x,u)\right) +i
\bigl{[}V_a(x,u) \stackrel{\ast}{,} V_b(x,u) \bigr{]}\label{Wab}
\ee
}

The action of the supersymmetric HS YM is proportional to
\be
{\frac 1{4g}}\langle\!\langle \ W^\alpha\ast W_\alpha  \vert _{\theta\theta}+ 
\bar W_{\dot\alpha}\ast \bar W^{\dot\alpha} \vert _{\bar\theta\bar\theta}\rangle\!\rangle\label{WWaction}
\ee 
It contains in particular the term $\langle\!\langle \ W_{ab}\ast W^{ab} \rangle\!\rangle$. With the redefinition $V_a(x,u)=- 2 h_a(x,u)$, we get {$W_{ab}= -4 G_{ab}$}, where $G_{ab}= \partial_a h_b-\partial_b h_a -i[h_a \stackrel{\ast}{,} h_b]$.   Then the action \eqref{WWaction} becomes             
\be
 \langle\!\langle -\frac 1{4g}G_{ab}\ast G^{ab} -i \bar \lambda \ast \bar \sigma_a \ED^a_\ast\lambda+ \frac 12 D\ast D\rangle\!\rangle\label{GGaction}
\ee
where $\ED_{a}^\ast \lambda(x,u) = \partial_a \lambda(x,u) -i[h_a(x,u) \stackrel{\ast}, \lambda(x,u)]$.

\section{Local Lorentz covariance}
\label{s:LLC}

To give way to a gravity interpretation we have to introduce local Lorentz invariance. To this end we use the teleparallel approach. Let us start from the definition of trivial frame. A trivial (inverse) frame
$e_a^\mu (x)$ is a frame that can be reduced to a Kronecker delta by means of a
local Lorentz transformation (LLT), i.e.  
such that there exists a (pseudo)orthogonal transformation $O_a{}^b(x)$ for
which
\be
O_a{}^b(x) e_b{}^\mu (x) = \delta_a^\mu\label{trivialframe}
\ee
As a consequence $e_b{}^\mu (x)$ contains only inertial (non-dynamical)
information.
A full gravitational (dynamical) frame is the sum of a trivial frame and
nontrivial piece
{
\be
\tilde{\tilde E}_a^\mu(x) = e_a{}^\mu (x)-\tilde{\tilde\chi}_a^\mu(x) \label{fullframe}
\ee
}
By means of a suitable LLT it can be cast in the form
\be
E_a^\mu(x) = \delta_a^\mu -\chi_a^\mu(x) \label{fullframetr}
\ee
This is the form one encounters in HS theories ($\chi_a{}^\mu(x)$ is the second component of  $h_a^\mu(x,u)$). But it should not 
be forgotten that the Kronecker delta represents a trivial frame. If we want to
recover local Lorentz covariance,
instead of $\partial_a=\delta_a^\mu \partial_\mu$ we must understand 
\be
\partial_a = e_a{}^\mu(x) \partial_\mu, \label{truepartiala}
\ee
where $e_a{}^\mu(x)$ is a trivial (or purely inertial) vielbein. In particular,
under an infinitesimal LLT, it transforms according to
\be
\delta_\Lambda  e_a{}^\mu(x) = \Lambda_a{}^b(x) e_b{}^\mu(x)\label{deltaLea}
\ee 

 A trivial connection (or inertial spin connection) is defined by
\be
\EA^a{}_{b\mu} = \left(O^{-1} (x) \partial_\mu O(x)\right)^a{}_b\label{teleA}
\ee
where $O(x)$ is a generic local (pseudo)orthogonal transformation (finite local 
Lorentz transformation).  As a consequence its curvature
vanishes
\be
\ER^a{}_{b\mu\nu} = \partial_\mu \EA^a{}_{b\nu}- \partial_\nu \EA^a{}_{b\mu} 
+ \EA^a{}_{c\mu}\EA^c{}_{b\nu}- \EA^a{}_{c\nu}\EA^c{}_{b\mu}=0\label{teleER}
 \ee
Let us recall that the space of connections is affine. We can obtain any
connection from a fixed one by adding to it adjoint-covariant tensors, i.e.
tensors that transform according to the adjoint representation. When the
spacetime is topologically trivial we can choose as origin of the affine space
the 0 connection. The latter is a particular member in the class of trivial
connections. This is done as follows. Suppose we start with the spin connection 
\eqref{teleA}. A Lorentz transformation  of a spin connection $ \EA_\mu =
\EA_\mu{}^{ab}\Sigma_{ab}$ is
\be
\EA_\mu(x) \rightarrow L(x) D_\mu L^{-1} (x)= L(x) (\partial_\mu + \EA_\mu)
L^{-1} (x)\label{LLL}
\ee
where $L(x)$ is a (finite) LLT. If we choose $L=O$ we get
\be
\EA_\mu(x) \rightarrow 0\label{Lfixing}
\ee
Choosing the zero spin connection amounts to fixing the local Lorenz gauge.

The connection $\EA_\mu$ contains only inertial and no gravitational
information. It will be referred to as the {\it inertial connection}. It is a
{\it non-dynamical} object (its content is pure gauge). The dynamical degrees of freedom will be
contained in the adjoint tensor to be added to $\EA_\mu$ in order to form a fully dynamical spin 
connection. $\EA_\mu$ is nevertheless a connection and it makes sense to define the inertial
covariant derivative
\be
{\cal D}_{\mu a}{}^b=\delta_a^b  \partial_\mu -\frac 12\EA_{\mu a}{}^b\label{inertialcovder}
\ee
which is Lorentz covariant and acts on tensor. On a two-spinor $\lambda_\alpha$ it takes the form 
\be
{\cal D}_{\mu\alpha}{}^\beta =\delta_\alpha{}^\beta \partial_\mu -\frac 12 \EA_\mu^{ab}\, \sigma_{ab\,\alpha}{}^\beta \label{inertialcovderlambda}
\ee
and for $\bar\lambda^{\dot\alpha}$
\be
{\cal D}_\mu{}^{\dot\alpha}{}_{\dot\beta} =\delta^{\dot\alpha}{}_{\dot\beta} \partial_\mu -\frac 12 {\EA_\mu^{ab }}\, \bar\sigma_{ab}{}^{\dot\alpha}{}_{\dot\beta}
\ee
In order to implement local Lorentz covariance we must replace ordinary partial derivatives $\partial_a$ with $e_a^\mu(x) {\cal D}_\mu$ everywhere, in particular in the definition of $D_a, \bar D_a$:

\be
D_\alpha &=& \frac {\partial}{\partial \theta^{\alpha}}+ i\sigma_{\alpha\dot\alpha}{}^\mu\bar\theta^{\dot\alpha} {\cal D}_\mu , \quad\quad \sigma^\mu(x)= \sigma^a\,e_a^\mu(x)\label{newDalpha}\\
\bar D^{\dot\alpha}&=&  -\frac {\partial}{\partial \bar\theta^{\dot\alpha}}- i \theta^\alpha\sigma_{\alpha\dot\alpha}{}^\mu {\cal D}_\mu
.\label{newDbaralpha}
\ee
Of course the new superspace covariant derivatives must satisfy the same relations as the old ones, i.e.
\be
&&[D_\alpha, D_\beta]=0,\quad\quad [\bar D_{\dot\alpha}, \bar D_{\dot\beta}]=0,\0\\
&&[D_\alpha,  \bar D_{\dot\beta}]= -2i \sigma_{\alpha\dot\beta}{}^\mu {\cal D}_\mu\label{DDbarDbarDnew}
\ee
One can verify that this is indeed the case since $[{\cal D}_\mu,{\cal D}_\nu]=0$ and ${\cal D}_\mu e_a^\nu(x)=0$.

The recipe for local Lorentz covariantization is:
\begin{enumerate}
\item replace any spacetime derivative, even in the $\ast$ product, with the inertial covariant derivative ${\cal D}$, 
\item interpret any flat index $_a$ attached to any object $O_a$ as $e_a^\mu(x) O_\mu$,
\item in any spacetime integrand insert $e^{-1}$.
\end{enumerate}

\vskip 1cm 

{At this point the stage is set for quantization. We do not intend to tackle this huge task here, but we would like to point out that the prospects in this sense are far from bleak. First of all we intend to consider BRST quantization. In \cite{HSYM} this has already been done for the non-supersymmetric version of the theories presented here and it proved to be straightforward. For the supersymmetric theory one has to  reconcile BRST with supersymmetry. This is indeed possible and the superfield formulation of N=1 supersymmetry (in 4d) is a privileged ground where this can be implemented, as was shown on ref.\cite{bonora2016}. This reference suggests that it might be a good idea to enlarge the supersymmetric superspace by adding additional anticommuting coordinates to the supersymmetric ones in order to represent the BRST symmetry.  Next, one has to define a concrete perturbative expansion by means of  (super) Feynman diagrams. In so doing, very much like in  \cite{HSYM}, one is bound to come across the problem of further gauge fixing in order to be able to invert the quadratic kinetic operator and write down propagators. Finally one will have to introduce (super)projectors in order to exclude negative norm states and guarantee unitarity and Lorentz covariance. There seem to be no obstacles in this direction. The real, still unexplored, problem with YM like theories is renormalization.  In this sense we expect supersymmetry may be of great  help. 
}

\section{Conclusion}

This concludes our construction of HS YM-like supersymmetric theories. Their action is  given by \eqref{GGaction}. This action is supersymmetric because it corresponds to \eqref{WWaction}, which is supersymmetric by construction, being expressed solely in terms of superfields. This action is also invariant under supergauge transformations, see (\ref{deltaVaxu},\ref{deltalambdaxu},\ref{deltaDxu}), which contain in particular diffeomorphisms. Therefore these theories admit a gravity interpretation, but, as formulated in \eqref{WWaction} or \eqref{GGaction}, they are not local Lorentz invariant. However in the previous section we have shown how to modify with formal steps the action in such a way as to implement local Lorentz invariance. The final action has therefore all the ingredients of a gravity action, being simultaneously supersymmetric. The original \eqref{WWaction} or \eqref{GGaction} are nothing but local Lorentz gauge fixed version of the latter. It is remarkable that the final theory may describe gravity in a supersymmetric form without resorting to supergravity.

\section*{Acknowledgements}
The authors would like to thank M. Cvitan, P. Dominis Prester, M. Pauli\v{s}i\'{c}, T. \v{S}temberga and I. Vukovi\'{c} for discussion and collaboration in the early stage of the project.
The research of S.G. has been supported by the Israel Science Foundation (ISF), \\ grant No. 244/17.

\end{document}